# Large volume ‘chunk’ lift out for 3D tomographic analysis using analytical plasma focussed ion beam – scanning electron microscopy

Ruth Birch[1*], Shuheng Li[1], Sharang Sharang[2], Warren J. Poole[1], Ben Britton[1]

1. Department of Materials Engineering, UBC, Vancouver
2. Tescan Brno, s.r.o., Czech Republic

*ruth.birch@ubc.ca

**Abstract**

Characterization of the structure and properties of materials in three dimensions, including grains and the residual pattern of deformation, provides necessary information required to guide materials design as well as support materials modelling efforts. In this work, we present an overview of site-specific large volume 'chunk' lift out and 3D serial sectioning of substantive volumes (e.g. 200 x 200 x 400 μm$^3$), where sectioning is optimized for 3D electron backscatter diffraction (EBSD) based crystallographic analysis, using a plasma (Xe) focussed ion beam scanning electron microscope (plasma FIB-SEM) equipped to perform EBSD using a 'static' configuration (i.e. slicing and EBSD-mapping are performed without moving the sample). This workflow is demonstrated through the 3D plasma FIB-SEM based EBSD analysis of an indent made within a polycrystal of pure magnesium. The lift out approach is suitable for a wide range of materials, and we offer a step-by-step guide within the present work to provide opportunity for others to more easily enter this field and collect valuable data.

**Graphical Abstract**

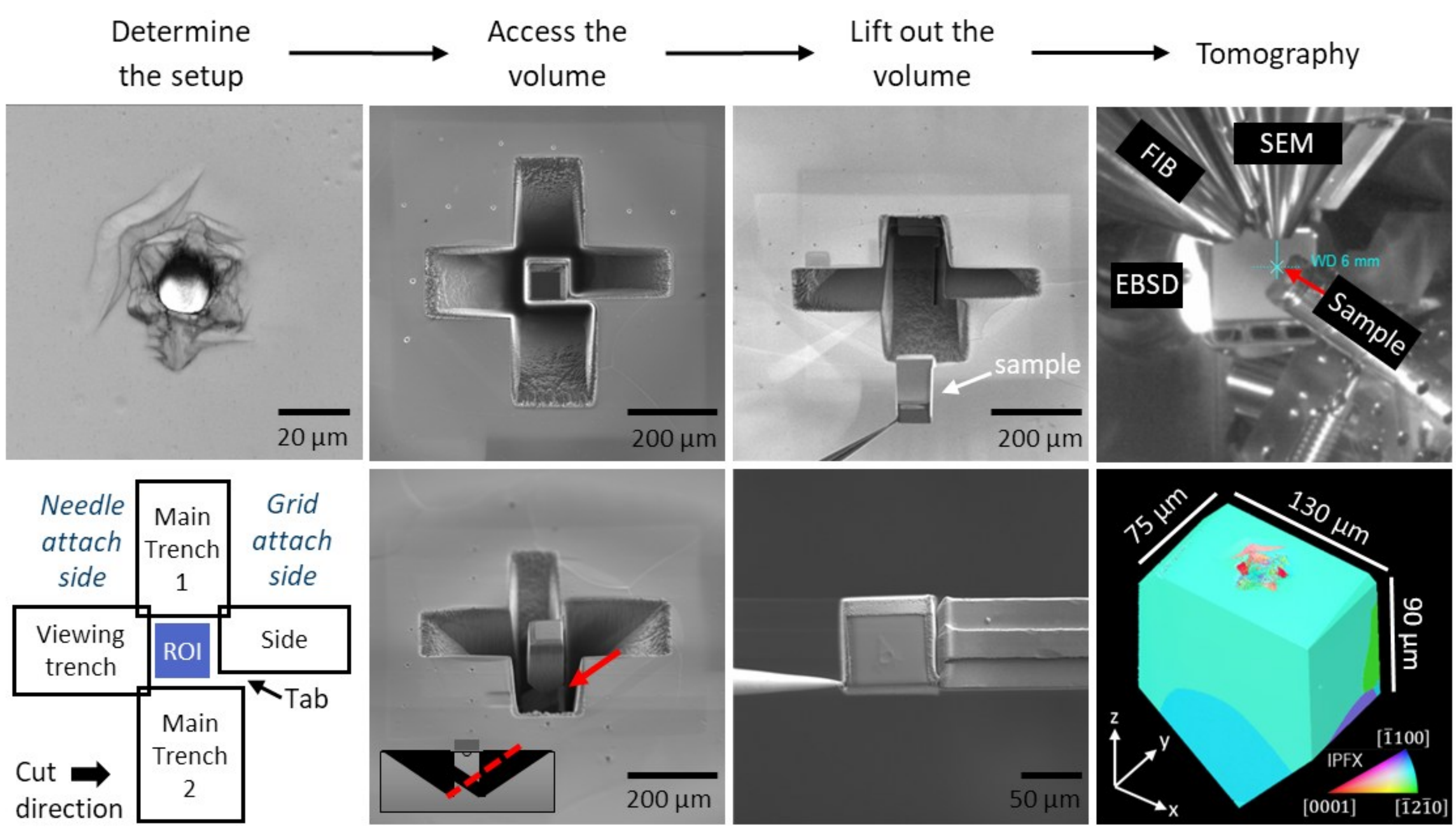

## 1 Introduction

The nature of most materials is that the 3D microstructure is important when using the material for important industrial applications. When only surface-based 2D characterization is used, the 3D microstructure is not properly accounted for, and scientific understanding can be limited or misconstrued due to the influence of the sub-surface volume. Despite this importance, access to 3D microstructural information is difficult to collect, and typically expensive, even if it contains useful information that typical 2D micrographs cannot easily capture or measure.

For example, 3D-microstructural analysis including both spatial and crystallographic information enables full 5 parameter grain boundary analysis [1], where 2D analysis cannot access the subsurface grain boundary trace of an individual boundary or statistical methods must be used to reconstruct the 3D distribution [2]. Another example is the analysis of 3D volumes around residual indentation impressions to show the full volume, rather than a signature of only the surface or one cross section [3]. In more complicated microstructures such as those found in steels, a full 3D analysis allows access to the full morphology of grains e.g. laths in martensitic steel [4] or around cracks [5], and is particularly useful for analysing additively manufactured materials that have anisotropic microstructures [6].

One increasingly popular method to capture and analyse the 3D microstructure is tomography using a focussed ion beam (FIB) and a scanning electron microscope (SEM). In these methods, the collection of a 3D tomography data set requires destruction of the volume of interest as each 'slice' of a 3D volume is removed in sequence and the new surface is analysed using SEM-based analytical techniques, including electron imaging (e.g. secondary or backscatter), chemical analysis with energy dispersive X-ray spectroscopy (EDS) (e.g. [7], [8]) or crystallographic analysis with electron backscatter diffraction (EBSD) (e.g.[6], [9]).

There are now commercial plasma focussed ion beam scanning electron microscopes (FIB-SEM) that can serially section through large volumes. Xe-ions are used as a popular plasma source as the focussed Xe-plasma typically has a much faster material removal rate than the Ga+ ions that are generated from traditional liquid metal sources [10]. The resolution of the data set is constrained by the slice thickness, where plasma FIB based methods typically provide a minimum slice thickness of ~10 nm, and the area studied per slice depends on the strategy for electron microscopy based mapping (i.e. the resolution and beam deflection over areas) which can extend >100 x 100 $\mu m^2$.

There are three types of 3D-EBSD setups used in a FIB-SEM:

(1) Tilt: the EBSD detector is located opposite to the FIB-column, and the sample is tilted between the slicing and EBSD mapping steps (e.g. [11]) ;

(2) Rotation: the EBSD detector is located below the FIB, with a rotation and stage tilt required to switch from the milling to mapping positions (e.g. [12], [13]);

(3) Static: the sample is located such that no movement is required (e.g. [14]).

The static EBSD setup is advantageous as it removes the potential for errors related to stage/sample movements. There is no additional time required for movement between slices reducing the overall time and allowing for greater throughput.

One challenge for analytical experiments, beyond simple imaging, is to ensure that the signal obtained from each new slice can be collected fully, i.e., there is a clear path for the exiting X-ray or electrons to reach the detector, and this challenge is magnified significantly in an EBSD-based experiment as typically the EBSD-camera requires access to diffracted electrons that can be deflected up to 50° away from the surface normal of the sample. To address this challenge, 3D experiments must either be performed towards the edge of a polished surface, use large volume trenches to allow the required exit angles (e.g. similar to the proposed workflow in [8] that was used for EDS-analysis) or alternatively, a site specific lift out technique can be used to extract a region of interest (ROI) from a bulk sample and place it on an analysis coupon, for subsequent plasma FIB-SEM based EBSD slice and view analysis.

Site specific lift outs are a common approach for very small volume analysis, e.g. atom probe tomography, where small volumes are lifted out from site specific areas and then shaped into needles before analysis [15]. Similarly, FIB-SEM setups are used for lift outs for (scanning) transmission electron microscope ((S)TEM) lamella preparation, where thin sections are cut from site specific regions and then thinned for (S)TEM analysis. Either as a standard lamella [16], [17], or in an alternative geometry, for example plan view [18].

For larger volumes, e.g. for 3D-EBSD analysis, a 'chunk' lift out can be used. Examples include a 'block' approach that has been used for 3D-EDS [19] and a similar approach for EBSD (described briefly in [20] and [21]). Chunk lift outs involve cutting out a large, cubic/cuboidal volume, from surface of a sample.  This volume can then be attached to a grid and used for tomography, in the same way as tomography at the edge of a sample. The advantage of chunk lift outs is the ability to be site specific and tailor the process (e.g. dimensions) to the region of interest, for example, to capture a specific region of an interface-/grain- or phase- boundary or the deformed volume surrounding a nanoindent impression.

In this paper, we provide a detailed description of a chunk lift out method to prepare a sample for 3D EBSD in a static setup which highlights the critical steps that can be optimized to maximise the value of the data that is ultimately collected. For an example case study, we briefly study a volume that surrounds a spherical indent made in a Mg polycrystal.

## 2 Methods

To demonstrate the approach, a site specific lift out of the residual impression associated with local deformation caused by nanoindentation of commercial pure magnesium is used. A site specific lift out is required for this analysis, as indentation was carried out within one grain of a polycrystalline sample and the indentation experiment was performed away from the edge to simplify the mechanical analysis.

The initial sample was as-cast commercially pure Mg as supplied by McMaster University. This sample was heat treated at 500 °C for 48 h with the sample in a steel bag and flowing argon to reduce oxidation. The sample was then mechanically ground to 2400 grit and diamond polished using 6µm, 3µm, 1µm and 0.02µm suspensions. This was followed by etching ($HNO_3$ 10% in volume + ethanol) for 3 mins. Optical microscopy was used to check the surface and average grain size followed by EBSD to determine whether there were suitable grain orientations in the sample. With suitable grain(s) identified, the sample was electropolished for 2 h at room temperature ($H_3PO_4$: ethanol=3:5) using a voltage of 2V with steel as the cathode. The sample was then cleaned in ethanol and dried using compressed air.

Nanoindentation was carried out on a Bruker TI Premier instrument using a spherical indenter tip and for the example dataset, a tip radius of 9.7 µm and max displacement of 5163 nm were used.

The lift out and subsequent analysis were performed using a TESCAN AMBER-X Xe-plasma focussed ion beam scanning electron microscope equipped with an Oxford Instruments Symmetry S2 EBSD detector. This detector is mounted on a port that enables use of the 'static' EBSD configuration, which is important for how the sample is finally mounted and analysed in 3D.

Preparation of the sample was enabled using a tungsten-based precursor for electron-beam and ion-beam deposition of protective layers and also joining the sample to the lift out needle and copper half grid, using an optiGIS gas injection system (GIS) and a TESCAN XYZ nanomanipulator system. The chunk lift out steps are summarised in Table 1, and each step is discussed in more detail in the next section. For the interested reader, a more detailed table of steps for the lift out is provided in the supplementary material (table S1) along with additional figures (S1 and S2).

*Table 1 - Beam conditions and stage tilt for chunk lift out steps*

| ID | Step | | Stage tilt (°) | Beam | Voltage (kV) | Current (nA) |
|---|---|---|---|---|---|---|
| 1 | Deposition (Tungsten) / Capping | e-beam | 0 | SEM | 5 | 10 |
| 2 | | i-beam | 55 | FIB | 15 | 120 |
| 3 | Trenching | main trenches | 55 | FIB | 30 | 1000 |
| 4 | | 'tidy up' | 55 | FIB | 30 | 300 |
| | | | 55 | FIB | 30 | 100 |
| 5 | Undercuts | undercut 1 | 0 | FIB | 30 | 300 |
| 6 | | undercut 2 | 0 | FIB | 30 | 300 |
| 7 | Lift out | needle attach | 0 | FIB | 30 | 250 |
| 8 | | cut tab | 0 | FIB | 30 | 300 |
| 9 | Attach | attach to grid | 0 | FIB | 15 | 120 |
| 10 | | needle cut | 0 | FIB | 30 | 100 |

For this configuration, the sample is mounted on a tall post which reduces the risk of collision with the EBSD detector during the 3D analysis (see Figure 1). The static configuration can be operated with the EBSD camera fully inserted throughout, however, in this experiment the detector was partially retracted by 50 mm while the FIB was removing each new surface layer to reduce re-deposition and damage to the phosphor screen.

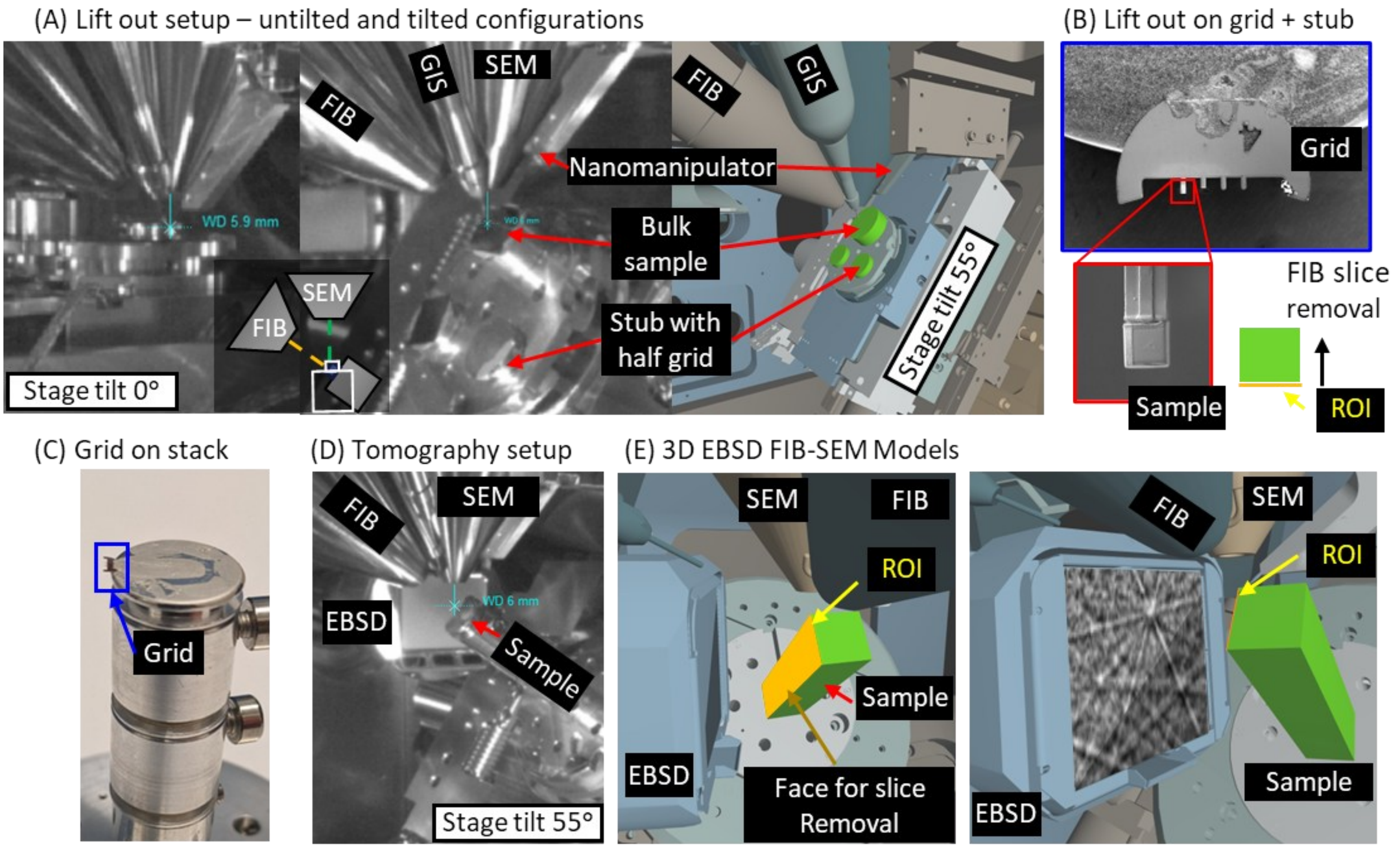

*Figure 1 – Region of interest (ROI) based lift out & 3D setup for focussed ion beam (FIB) and scanning electron microscopy (SEM) based electron backscatter diffraction (EBSD) analysis (A) chamber setup for lift out – chamber view (0 and 55° tilted) and model (55° tilted); (B) top-down view of the sample as mounted on the copper half grid prior to 3D analysis. (C) macroscopic view of the grid as mounted on a stub and tall holder; (D) chamber set up for tomograph; (E) chamber models showing the static EBSD set up. In brief, in the static set up the FIB removes one (yellow) layer at a time from the 3D volume, and the SEM is used to produce an EBSD-based microstructure map of each layer without moving the sample between slice and view steps.*

In this static setup for 3D EBSD experiments, the cut surface is facing the EBSD detector at a 70° tilt, whilst the edge being cut is facing the FIB beam, as shown in Figure 1 & Figure 2. In this geometry, no stage movement is needed during the experiment which speeds up the 3D analysis and reduces artifacts. The FIB-SEM coincident point is 6 mm for this system.

For the static setup, we need to define the axis systems, as shown in Figure 2, where all axis systems defined are right handed. To relate the axis systems, we can start with EBSD pattern (p) to SEM (s): $Y_p \parallel Z_s$ and $X_p$ is along $-X_s$ but rotated by 14.7°. EBSD pattern to sample conventions verified following Britton et al. [22]. For imaging and analytical (i), i.e. SE and EBSD, there is a scan rotation of 284.7 °. This axis system is related to the SEM axis using: $Z_i \parallel Z_s$ and $X_i$ is along $X_s$ but rotated 14.7°. It is related to the tomography sample axis system (t) by: $Z_i \parallel -Y_t$ and $X_i$ to $X_t$ is ~55° rotation about $Z_i$.

In this case, the tomography and FIB imaging axis systems are identical, noting that the FIB images have a scan rotation of 295°. SEM to FIB is therefore the same as imaging to tomography.

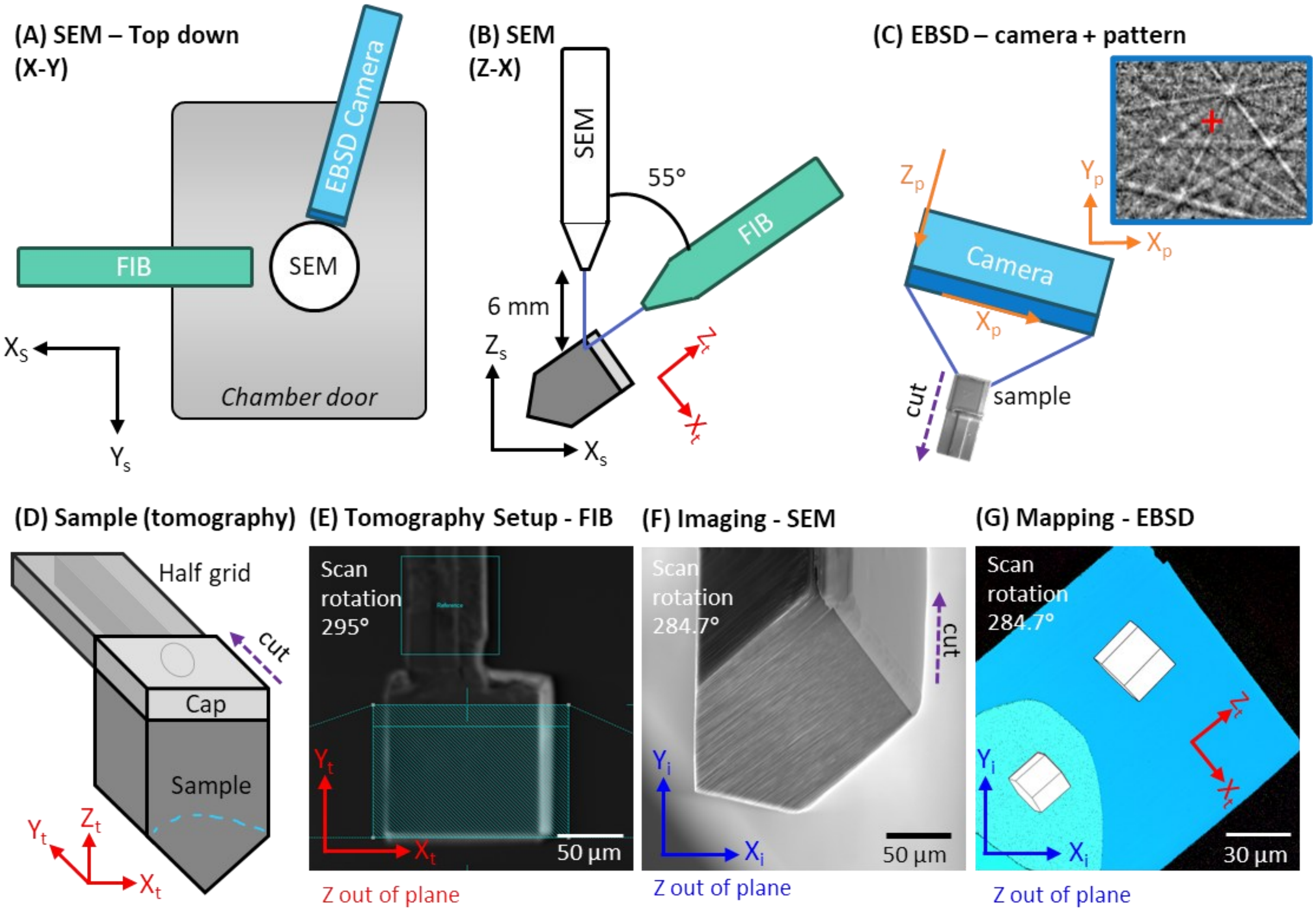

*Figure 2 - Diagrams showing axis systems: schematic of SEM layout in (A) X-Y and (B) Z-X planes; (C) EBSD camera and pattern; (D) tomography sample diagram; (E) FIB tomography setup; (F) SEM imaging; (G) EBSD mapping. Note: all axis systems are right handed. Purple dashed arrows indicate the slicing (cut) direction.*

For the tomography, the EBSD for each slice was collected using 20kV 10nA for an area of 150 x 150 µm$^2$ at a step size of 0.15 µm. Slices of 0.15 µm in thickness were cut using the FIB at 30kV 100nA, thus giving a voxel size of 0.15 µm$^3$. Each slice took 1 min 42s to cut and ~6 mins to map. The complete dataset (150 x 120 x 101.5 µm$^3$) took just over 4.5 days to collect, as there is extra overhead associated with the tomography data (e.g. overview imaging and alignment steps, for a total of ~20% of the total experiment time). After the data was collected, the data was exported from AZtec, loaded and rotated in MATLAB using MTEX version 5.10.2 [23] and visualised in Dragonfly 2024.1 [24] .

## 3 Chunk lift out (Results)

For this manuscript, a series of nanoindentations were made in magnesium. One of these indents was selected and subsequently lifted out and analysed, with the aim of capturing the complete indent using EBSD. The indentation was carried out in a central grain on the Mg sample to provide freedom about which grain was selected and to simplify the mechanical analysis (i.e. away from the sample edge). This means that a lift out was required to extract the residual impression around the indent impression and perform volume-based analysis. This provides a case study of the lift out and tomography process, described in detail below.

### *3.1* Determining the setup

Prior to conducting the tomography experiment, appropriate FIB conditions were determined for both the lift out and the tomography experiment using a series of preliminary trials. The aim was to choose FIB parameters that would (1) mill the most material as quickly as possible (e.g. trenching), reducing the overall time required and (2) reduce surface damage, redeposition, and any ion-induced phase transformations during preparation and/or polishing. In Mg for example, material near the Xe-irradiated volume (and sputter deposition) can twin extensively when high FIB currents are used and this is both a problem for trenching and the tomography experiment. Initial testing showed that FIB currents above 100 nA during tomography would transform any (indentation induced) twinning around the indent during the polishing and make any comparison with the 3D tomograph and the initial experiment difficult. In practice, a higher current was determined to be satisfactory for trenching (i.e. far away from the final volume) to provide access to the final volume, as the FIB-induced twinning (see Figure 3 (D-ii)) did not impact the final volume analysed with 3D EBSD.

EBSD conditions were also tested on an example cross section face – the aims here were (1) to check that the FIB polished surface was of sufficient quality to collect EBSD data (2) to find appropriate mapping conditions (area, step size, acquisition speed) to reliably map each slice of the target volume and (3) to approximate the time that would be required per slice.

Both optimisation steps were carried out on a test indent in a grain of similar orientation to the final indent to avoid orientation related differences in FIB milling. The target indent was then decided based on some initial surface imaging and EBSD and then the time requirement for the tomography approximated based on (1) the size of the indent and therefore target volume (2) milling time per slice with the optimum FIB conditions and (3) EBSD (& imaging) time per slice. High quality EBSD of (the surface of) the selected indent was collected prior to the lift out to allow comparison with the 3D reconstruction after the tomography experiment.

### *3.2* Accessing the volume

The first part of the lift out process, accessing the volume, consisted of surface protection, capping, trenching and undercuts, with the aim of creating the final volume that will be lifted out and attached to the grid. It was mostly carried out at 0° stage tilt, with some of the initial steps at 55° stage tilt so the top surface of the sample was perpendicular to the FIB beam. The sample and stub with half grid were both in the chamber for the entire lift out process, located in opposite positions on the stage carousel to minimise potential collision issues.

After some initial imaging of the top surface of the indent, a relatively thin e-beam deposition layer was deposited to protect the surface of the indent. In the example shown in Figure 3 (B), a 1 µm thick e-beam layer of tungsten was deposited over an area of 80 x 80 $\mu m^2$ using a 5 kV-10 nA e-beam at 0° stage tilt. The deposition was checked using backscatter imaging, as the tungsten deposition layer appears brighter than the surrounding region.

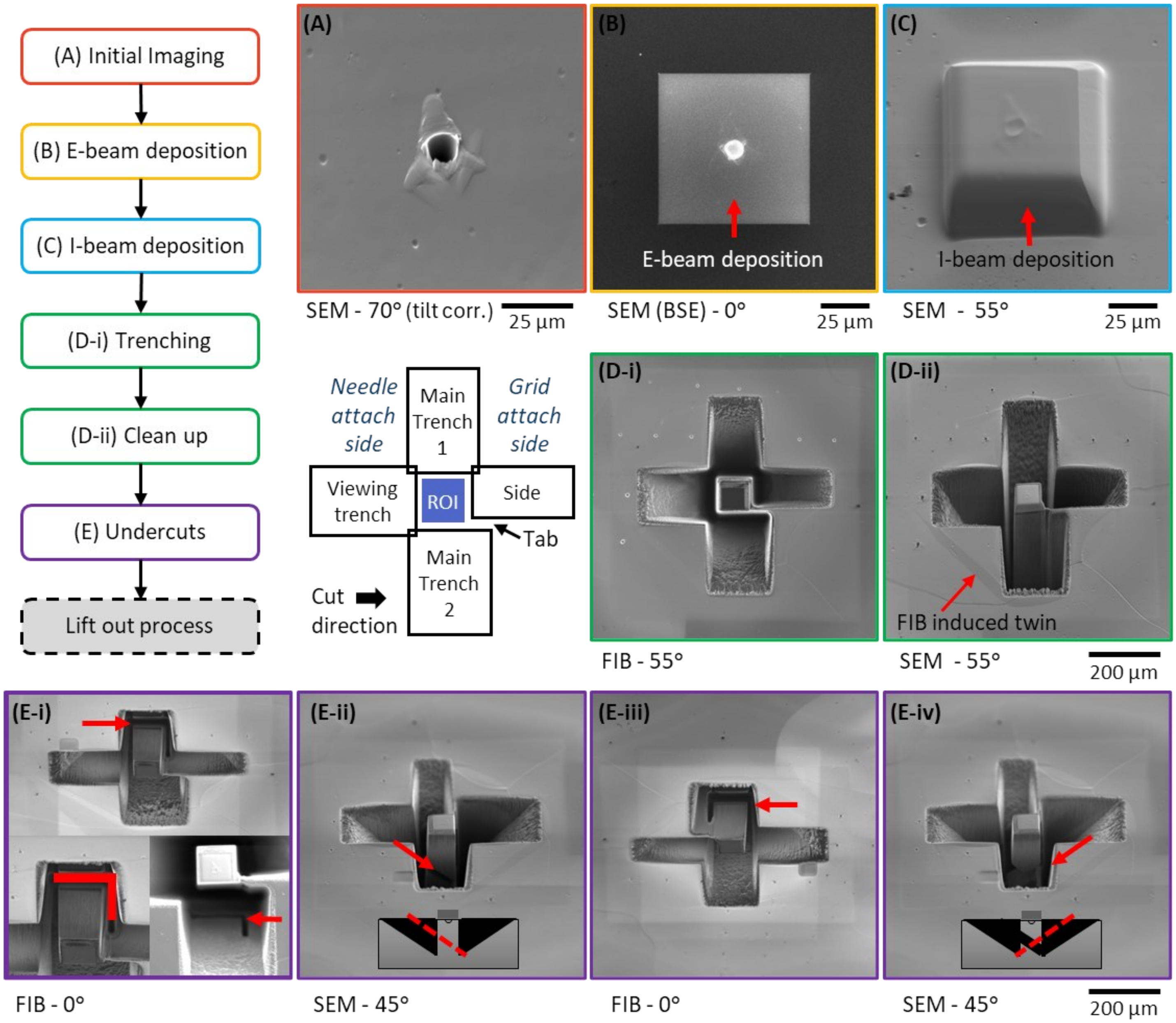

*Figure 3 – Initial process. (A) SEM view of the region of interest; (B) Surface protection using e-beam deposition – the W is brighter in backscatter; (C) Capping using i-beam deposition; Trenching: (D-i) post clean up FIB view and (D-ii) SEM view with FIB induced twin indicated; Undercuts: FIB views showing 'j-cuts' (E-I & E-iii) with 180° rotation between – (E-i) includes a zoom in of the cut (bottom left) and a SEM view (bottom right) of the cut once complete, which can be seen as indicated by the red arrow; and SEM views into the viewing trench showing the undercut progress (E-ii & E-iv).*

Next, a capping layer (i-beam deposition) was added to form a uniform layer and to minimise curtaining issues caused by uneven surfaces, differences in phase or orientation related differences in polishing rate. This step was carried out at 55° stage tilt (perpendicular to the FIB) to improve the accuracy of the deposition area. A small rotation was required when switching between 0 and 55° (SEM & FIB imaging) to keep the deposition patterns aligned. The same 80 x 80 $\mu m^2$ region was targeted, using a FIB-beam of 15 kV-120 nA. The cap thickness at this point was measured to be 32 µm.

This was followed by trenching, the longest part of this initial process, also carried out at 55° stage tilt. The pattern we used for trenching consisted of two main trenches for cutting the volume and two side trenches, one of which was used as a viewing trench for the undercuts, as shown in Figure 3 (E-ii & E-iv). The size and depth of the main trenches were chosen to ensure that the undercuts (carried out at 0° stage tilt) could be made. We used a FIB-beam of 30 kV-1 µA to cut the 4 trenches, with the size approximately 150 µm deep and 220 µm long. Trenching was followed by a series of polishing steps using a FIB-beam of 30kV-300 nA and then 30kV-100nA sequentially to 'square off' the volume after trenching, by polishing off the angles left by the beam tails and redeposition. Figure 3 (D-i) shows the pre- clean up FIB image and Figure 3 (D-ii) showing the post-clean up SEM view with a FIB induced twin indicated.

From this point on, everything was carried out at 0° stage tilt. The final step in creating the volume to lift out was the undercuts, where the final volume was cut, leaving just a small tab of material holding it in position. The incident angle of the FIB with the stage at 0° tilt creates a characteristic v shape at the bottom of the lift out and the depth at which the cut is made (with respect to the top of the trench) determines the length of the straight part of the sides. The undercuts used consisted of a long trench parallel to the edge of the cap, and a second cut perpendicular to this that went from the end of the initial cut to 2/3 of the way to the top of the sample as shown in Figure 3 (E-I to E-iv).

The undercuts were carried out at a FIB-beam of 30kV 300nA, with the cuts being stopped once there was evidence that the cut was complete all the way along. By optimising brightness & contrast, it was possible to see into the trench and observe the cutting line as the cut was made i.e. as it cuts through, it cuts a line in the wall of the trench – when this is a complete line, the cut is likely complete, as shown in the bottom right SEM image in Figure 3 (E-i). The viewing trench was used to check the cuts, between sides and at the end - by rotating the stage by 90° and tilting for a better view into the trench (e.g. to 45°), the undercut(s) could be viewed, as shown in Figure 3 (E-ii & E-iv). Wider cuts for the trench parallel to the edge of the cap were chosen to avoid issues with tapering due to redeposition, particularly during the undercut on the second side, that could have led to an incomplete cut and issues with lifting out the volume. After cutting the first side, a 180° stage rotation was used to move to the opposite side and complete the second set of undercuts, forming the final volume to lift out.

### *3.3* **Lifting out the target volume**

Next, the volume was lifted out and attached to the grid, and this is summarised in Figure 4 (lift out) and Figure 5 (attach to grid). To maximise success here, the sample and a stub with an attached half grid are loaded together into the chamber, as shown in Figure 1 (A).

In our microscope, the lift out needle approaches from the left of the screen (Figure 4A) and this means that the sample (and later the grid) must be aligned to position the edge of the sample parallel to the screen axis. The needle is inserted, together with the GIS, and the needle and sample are moved close together using both the SEM and FIB imaging to check alignment. W-deposition is used to weld the volume to the needle (Figure 4B), using a FIB-beam of 30kV-250pA and over an area of approximately 7 x 10 $\mu m^2$. With the needle attached, the tab was cut using a thin rectangular box at 30kV 300nA, running until the cut had been made, in this case, 3-4 mins. To ensure that the sample would align well with the half grid, the cut tab region was polished back to a flat edge to remove additional material left after the cut using 30kV 100nA (Figure 4 (D) shows post clean up). The sample was then lifted out on the lift out needle and moved out of the way whilst the stage was rotated to the grid.

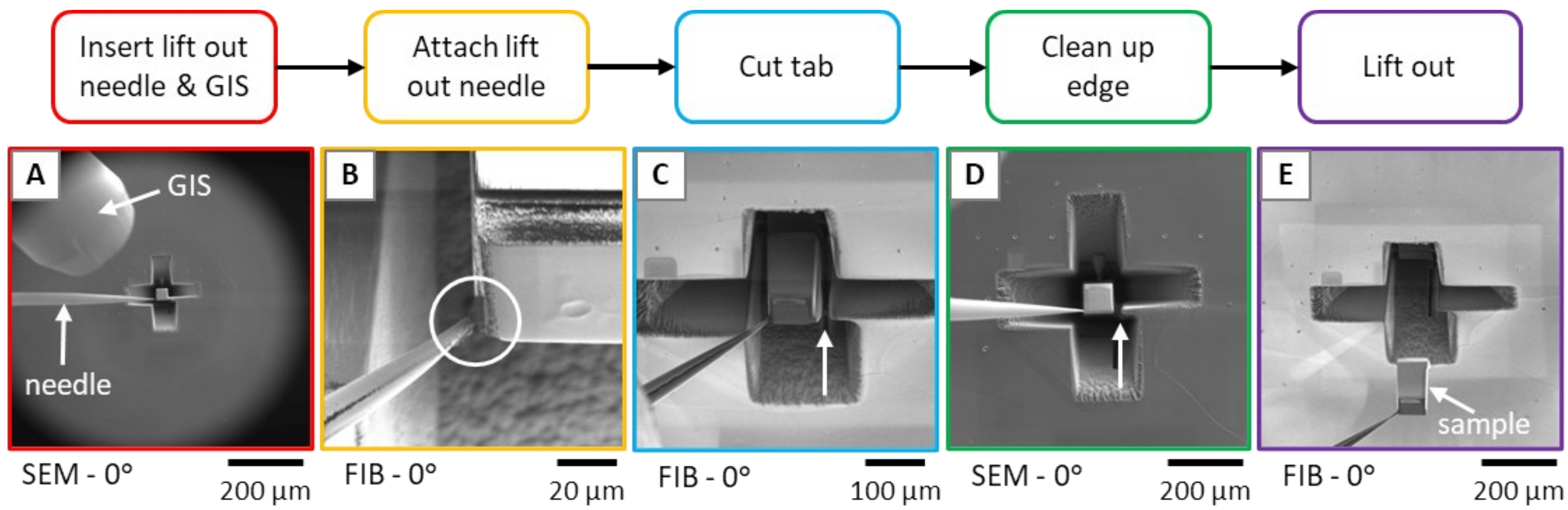

*Figure 4 – Key steps for the lift out process. (A) insert the lift out needle and gas injection system (GIS); (B) attach the lift out needle to the sample; (C) cut the tab; (D) clean up the edge of the lift out volume and (E) lift the sample out.*

Next, the Mg-sample was moved out of the way and the half grid was positioned into the field of view ready for attachment of the 3D volume (steps are shown in Figure 5). The half grid was positioned such that the flattened edge of the grid would be parallel with the edge of the sample volume (see Figure 5 (A)), and ensuring that it was possible to closely view both the end of the post (for deposition and attachment) and opposite side of the sample (to cut off the needle) as once the sample is attached to the grid, movement is limited. The sample was then slowly moved into position, aligned with the half grid post and with the top surface level with the top of the half grid. To avoid potential movement and collision issues when the GIS was re-inserted, the sample was moved a small amount away from the grid during insertion. The sample was then moved to its final position for attachment, aligned with the grid, but leaving a very small gap to make it easier to see the attachment progress. A large rectangular FIB-beam deposition box (15kV-120nA) was used to attach the sample to the grid as shown in Figure 5 (C).

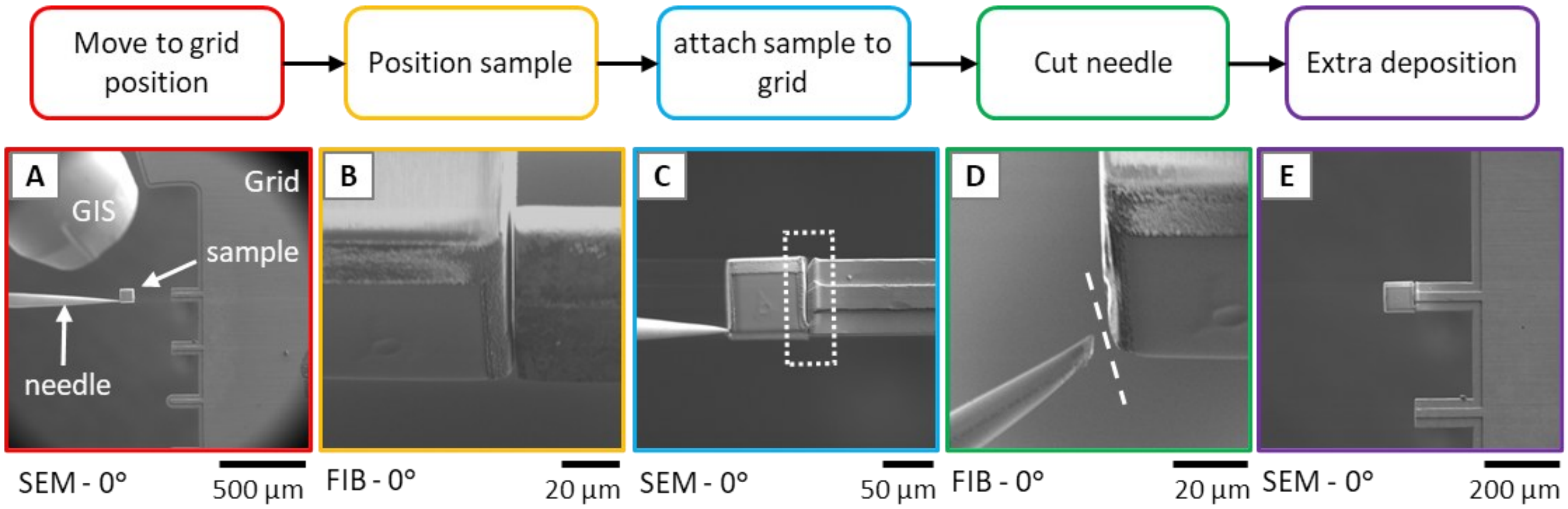

*Figure 5 – Key steps for attaching the lift out volume to the half grid post. (A) moving the sample into position – FIB view; (B) moving the sample into position – SEM view; (C) attach sample to grid; (D) cut the lift out needle free and (E) extra deposition when the sample is attached and free.*

Once the sample was satisfactorily attached to the half grid, the needle was carefully cut off. Note that in the geometry used for this example, the face the needle was attached to forms the front face of the volume and so damage to the surface would have needed to be cleaned up prior to use for tomography. In this case, the needle was cut using a thin rectangular box at 30kV 100nA (Figure 5 (D)). With the sample now free of the lift out needle and attached to the grid, further FIB-beam deposition was used to strengthen the weld, this time with the stage tilted to 55° for improved deposition.

The final preparation steps (also at 55° stage tilt) prior to the tomography experiment were (1) to add an alignment marker – here we cut a hole in the post (using a FIB beam burn) to use as a simple but effective fiducial marker that can be used for image registration during the 3D tomography experiment; (2) to clean up the front face, removing redeposition and ensuring a surface suitable for EBSD – clean-up was done using a FIB-beam of 30kV-100nA (this is the same conditions that would be used for the tomography).

### *3.4* Setting up the tomography experiment

The stub with the half grid on was then re-mounted onto a taller holder (as shown in Figure 1) which facilitates access to the static EBSD geometry, and the bulk sample removed from the chamber. In this static configuration for an AMBER-X instrument, the EBSD camera is mounted at an offset angle from the tilt axis of the microscope and so alignment of the 3D volume requires specific rotation and tilts. The sample was positioned for the 3D experiment by rotating to face the EBSD detector and tilting to 55° stage tilt so that the edge to cut is perpendicular to the FIB column. These frames of reference are described in Figure 1 and Figure 2. Software control of the tomography is a two-step process: (1) the AMBER-X control software (TESCAN ESSENCE) being used to define the cut volume and the analysis volumes, and fiducial based drift-correction/realignment; (2) the Oxford Instruments software is used to define the EBSD mapping parameters for each slice. Once these are established, the TESCAN ESSENCE tomography software will cut a slice, perform minor realignment to account for drift (i.e. using beam shifts of the SEM) and then ask the EBSD software to map, and this is repeated many times to generate the 3D volume.

For the tomography step, the cut volume (volume to be sliced) was chosen to be larger than the sample volume - slightly wider than the lift out volume to reduce any polishing artifacts at the edges and cover the whole width if the sides were not exactly parallel after the lift out process, and deep enough that the whole side is evenly removed. The start point of the box was placed outside the front edge of the lift out volume to avoid creating a shelf by starting to cut inside the volume instead of the front face. The FIB polishing parameters and the slice thickness were also defined as part of this setup, with the slice thickness chosen to give square voxels in the EBSD (i.e. EBSD step size = slice thickness).

The analysis volumes were defined by setting up the imaging and analytical data collection. The EBSD for each slice in our setup is square and is collected such that the top of the EBSD map is approximately 55° to the sample surface. The EBSD mapping area was chosen to contain the sample surface and a reasonable depth below the indent, not the entire lift out volume in this case. A second analysis volume was created by setting up collection of an overview image for each slice to monitor how the polishing is progressing and for troubleshooting if there was an issue (the resolution of this imaging was not chosen to be the same as the EBSD mapping to minimise collection time).

### 3.5 Example Dataset

In this section, we show an example dataset where the chunk lift out method described was used to cut out and map an indent in magnesium. The analysis volume of interest was 150 x 150 x 85 $\mu m^3$, with a voxel size of $(150 \text{ nm})^3$. A total of 569 slices were collected but a subset of 500 slices (75 µm) are shown in the figures below. The EBSD data was rotated (orientation & spatial) to keep a consistent axis system, as described in Figure 2 & Figure 6.

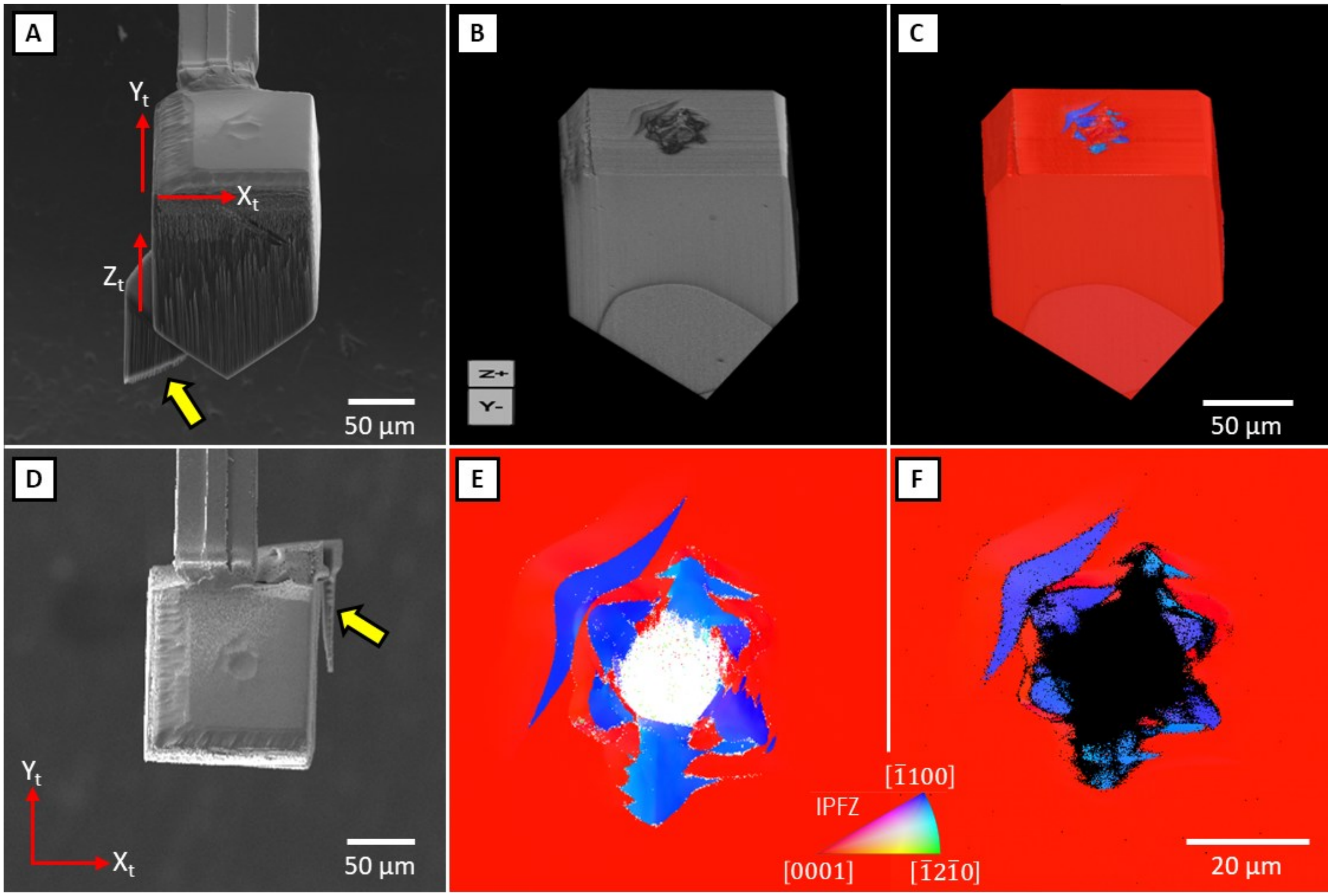

*Figure 6 – Mg indent lift out comparisons. Top row: similar 3D views of (A) the complete lift out volume prior to the experiment (SE); (B) analysis volume - band contrast; (C) analysis volume – IPFZ; Bottom row: top down views: (D) the lift out volume prior to the experiment (SE); (E) Top surface EBSD prior to the experiment- IPFZ; (F) 3D volume top surface EBSD data – IPFZ. Data has been rotated to have a consistent axis system.*

Figure 6 (A) and (D) shows the lifted-out volume prior to cleaning up the front face and starting the tomography. As can be seen (yellow arrows), the volume had a piece of debris from the lift out process attached (by Van der Waals) at the back of the sample. It was not possible to remove but did not interfere with the tomography. Pieces like this can be minimised by using wider cuts for the undercuts.

The reconstructed 3D volume agrees well with the SEM images of the initial volume, as can be seen by comparing the 3D renderings of band contrast (Figure 6 (B)) and EBSD data (Figure 6 (C)) to the initial volume (Figure 6 (A)). Comparison of the top surface EBSD before (Figure 6 (E)) and the reconstructed surface (Figure 6 (F)) also agree well. There is a slight rotation in both Z and X resulting from the lift out process which could be addressed through further rotations if direct comparison were required. Importantly, the twins remain throughout the tomography experiment.

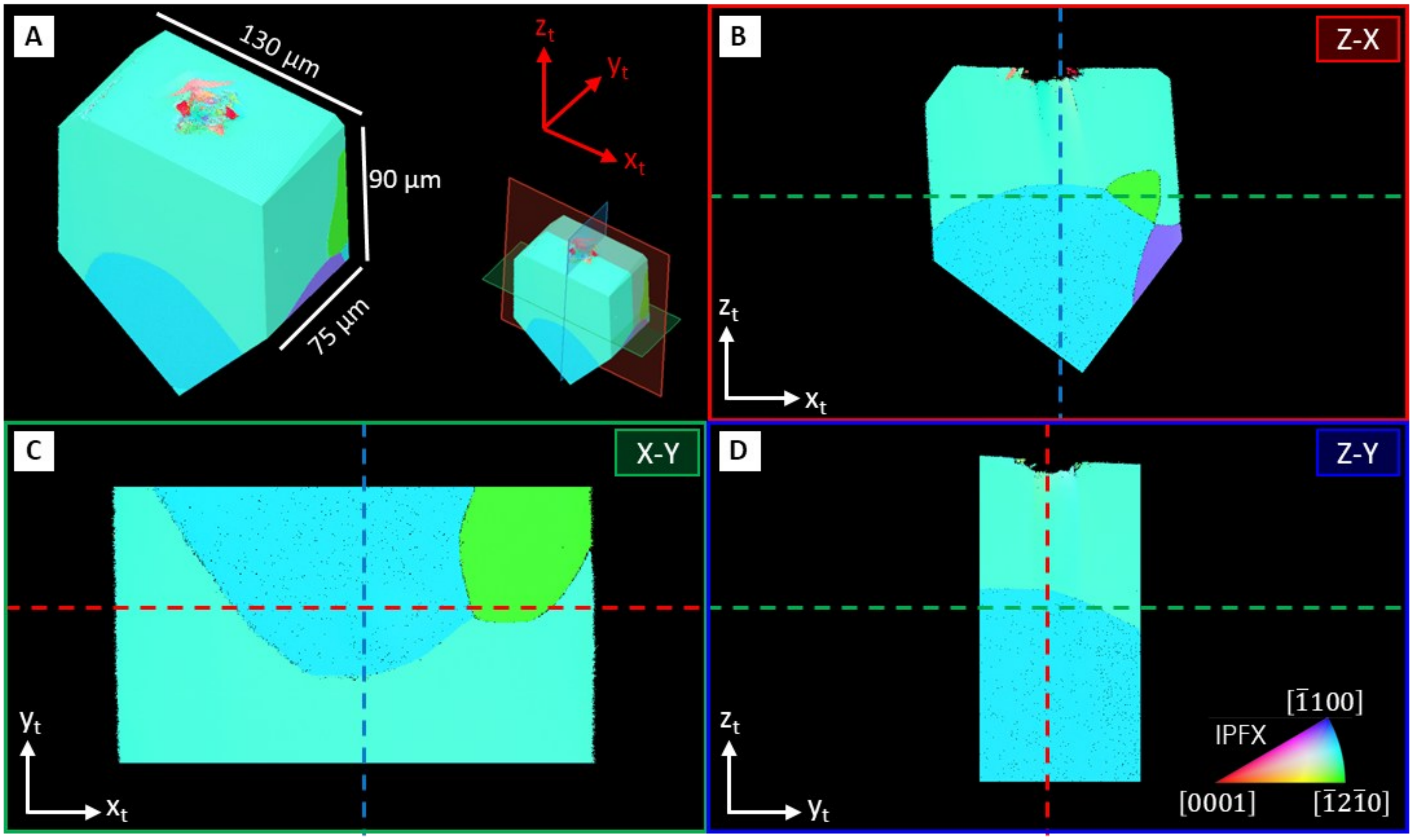

*Figure 7 – Cut-through views of 3D volume (IPFX colouring). (A) complete volume with measurements; (B) Z-X plane; (C) X-Y plane; (D) Z-Y plane. Axis systems follow conventions defined in an earlier figure.*

Figure 7 shows the 3D dataset using IPFX colouring, using a 3D view and a cut through on each plane. The first thing we can observe is that the whole of the indent appears to have been captured, with some surrounding (non-indented) volume as well. As the nanoindentation experiment had a goal of testing the properties of a <c> axis aligned grain, for the nanoindentation analysis it is unfortunate that there are sub-surface grains which likely impact some of the indentation data, especially for the greater depths of indentation. This issue does however highlight one of the useful things about 3D experiments – seeing the whole volume in detail. In this 3D volume, there are 4 grains identified and the grain boundaries (and triple junctions etc.) are clearly revealed in the resulting volume maps. Analysis of the deformed volume around the indent is out of the scope of the present work, as we hope to present in much more detail within a later publication.

## 4 Discussion

The manuscript has presented a detailed step-by-step approach for static 3D EBSD lift out and analysis, and a case study in Mg. The approach recovers the volume reasonably well, due the presence of the capping layer which protects the near-surface region. There is a slight misorientation between the original surface map and the 3D volume, and this orientation mismatch is likely related to the challenges associated with the absolute orientation of EBSD data (which typically has a ~2 degree uncertainty [25] as well as precise alignment of the sample and copper lift out grid during lift out. This misalignment could also arise at a number of other places within the lift out process: (1) when cutting and polishing the sample, beam tails and/or redeposition can cause a slope in any of the sides – careful clean-up of the sides can improve this, but there is a balance of time and effort and it is easy to reduce the target volume if not done carefully. (2) Rotation of the lift out volume on the needle can occur, particularly for larger lift outs. Keeping the sample on the lift out needle for the minimum amount of time can help with this, and for larger samples, locating the needle at the midpoint of the side should help with minimising rotation. (3) When attaching the sample to the grid, the W-based welding step can cause the volume to tilt slightly upwards as the gap is closed and the two sides are pulled closer together. (4) Alignment of the sample at the start of the tomography, i.e. if the edge of sample and the tomography box (cut line) are not parallel, this can also cause a skew in the final reconstruction. A useful feature of lift outs is that it is significantly easier to determine where the sample volume starts as there is no material remaining below and in front of the start point so the edge is easier to determine, due to the change in contrast (sample vs. free space), even if using a large current.

It may be possible to improve the recovery of the surface layer by adapting the method proposed here. One way would be to provide improved surface protection, either using a thicker layer of tungsten, or by filling the indent with deposition prior to capping it. Both routes may provide an improvement in the polishing at the surface by increasing the uniform layer for cutting and, in the latter case, allowing a uniform cutting rate across the sample. An alternative route, in the case where the shape is recovered well but the EBSD data is poor in the surface region (e.g. due to strain at the surface), is to pattern match the EBSD data. We have seen good improvement of the data recovered using this route, but it requires significant post processing time (e.g. a similar dataset took ~3 weeks of 24/7 processing) and requires the patterns to be saved with the dataset. In all likelihood, the 'best' method to improve surface recovery will likely be a balance of these methods.

The Xe-FIB-SEM and static 3D-EBSD-lift out approach presented here can be compared with the literature:

(1) The 'block' EDS approach of Schaffer et al. [19] is similar to the EBSD approach here, however in the present work the exit angle of the diffraction pattern is large compared to the solid angle of the EDS detector. Furthermore, the crystal quality has to be high so that the EBSD patterns can be reliably indexed [10]. This required tailoring of the FIB-slicing settings, as well as the mounting of the lifted out sample.
(2) Static EBSD can also be performed with a Ga-FIB, provided the chamber is configured for this approach (e.g. Guyon et al. [14]). Many of the ideas within this Xe-FIB based approach can be echoed in a Ga-FIB-SEM, with care to address any Ga-related artifacts (e.g. implantation, alloying, or grain boundary embrittlement). The size of the Xe-lift out is often much larger than available within a Ga-based approach, and this can impact how the volume is isolated for tomographic analysis.
(3) For other EBSD configurations (e.g. rotate and tilt [12], [13], or tilt [11] ), the large chunk lift out approach could be performed with modifications to the final geometry so that the isolated volume is presented well for FIB slicing and EBSD analysis. In the static configuration, time is saved when the geometry is optimized for high quality EBSD pattern capture with no sample movement between slices.

Inherently, as compared to 2D surface analysis, 3D experiments cost significant resource, in instrument time (preparation, lift out and tomography), processing time, data storage and human analysis. We have found that the development of optimized standard routes can significantly improve the reliability of an experiment and reduce costs (e.g. we can confidently use the same data processing pipeline). However, there will be some constraints that limit a 3D experiment, as for example in this case, we found we were limited in the FIB current we could use to polish the sample (affecting the cutting time) if we wanted to recover the (indentation induced) twins, as higher FIB currents would transform this surface region. Furthermore, initial work was done to check that the EBSD mapping experiment (per slice) could be useful for the analysis that we wished to undertake, and so the EBSD map size (analysis volume) was also determined in advance, a careful balance between covering a large enough volume around and below the indent, map quality (step size, exposure time, SEM conditions) and the time available. Depending on the experiment aim, different compromises can be made e.g. if the plan is to pattern match the data, then the EBSD data collection can be optimised for speed, which could increase throughput and allow a larger overall volume.

It is worth noting that data storage is also a significant consideration for 3D experiments both for the data collection computer and for data processing, and it is very tempting to 'over collect' information (as the SEM is equipped with multiple different sensors/detectors) especially for a destructive tomography experiment. This motivates the development of smart processing strategies that enable initial go/no go decisions for different analysis routes, as well as identifying what types of data reduction are possible. In the case of EBSD experiments for example, saving of every single pattern can have value, but may significantly increase the size of the data volume that is collected (especially if pattern matching is to be used later to improve orientation precision, or as noted, recovery of poorer quality information near the surface or interfaces). As an example, for the example shown, 1 slice of EBSD data with the EBSD patterns saved is 20 GB on the disk, but only 0.7 GB without these patterns saved. For the 3D tomography experiment, this rapidly can cause issues as now a 'simple' (and for a plasma FIB, quite small) 3D volume of 100 slices could now be a 2 TB data set which takes time to move from instrument to post processing computer and to storage, and also 'on microscope' issues may related to optimizing the configuration of hard disk systems which were simply not designed for a 3D tomography experiment (e.g. upgrading a SDD storage properly gets expensive very quickly).

## 5 Conclusion

We have detailed a chunk lift out process for getting large volume site specific lift outs for use in a static tomography setup. This process has 4 main stages: determining the setup, accessing the volume, lifting out the volume and tomography. We have demonstrated this method using an indent in polycrystalline Mg. Here, using the chunk lift out method allowed us to be site specific, targeting a specific grain orientation and avoiding complications in the mechanical analysis associated with indentation at the edge of a sample. We show we can lift out the entire indent that results in good recovery of the surface after reconstructing the data. This method also revealed the presence and shape of subsurface grains which would not have been known using 2D methods. The lift out approach is suitable for a wider range of materials, and we hope that the step-by-step guide within the present work provides opportunities to inspire others to more easily enter this field and collect valuable data.

## 6 Acknowledgements

The AMBER-X pFIB-SEM was funded by the BCKDF and CFI-IF (#39798: AM+). WP and SL acknowledge funding from the WP's Canada Research Chair in Through Process Modelling of Advanced Structural Materials. We acknowledge funding from NSERC Discovery Grant program (RGPIN-2025-04768, WP and RGPIN-2022-04762, TBB). We also would like to thank Colton Whittaker (Oxford Instruments) for helpful discussions and technical support.

## 7 CRediT statement

Ruth Birch – investigation, methodology, writing – original draft, writing - review & editing; Shuheng Li: investigation, writing - review & editing; Sharang Sharang; methodology, writing - review & editing; Warren Poole – supervision, funding acquisition, writing - review & editing; Ben Britton – supervision, funding acquisition, writing – original draft, writing - review & editing.

## 8 Data statement

High quality figures are available at 10.6084/m9.figshare.29373965.

## 9 Conflict of Interest

Sharang Sharang is an employee of TESCAN, who make and distribute the AMBER-X plasma FIB-SEM used in this work.

## 10 Supplemental

*Table S1 - Summary of steps used for chunk lift out*

| ID | Step | | Stage tilt (°) | Beam | Voltage (kV) | Current (nA) | Area (µm) | Depth (µm) | Duration |
|---|---|---|---|---|---|---|---|---|---|
| 1 | Deposition | e-beam | 0 | SEM | 5 | 10 | 80 x 80 | 1 | 35m 43s |
| 2 | | i-beam | 55 | FIB | 15 | 120 | 80 x 80 | 35 | 1h 20m |
| 3 | Trenching | Main trenches | 55 | FIB | 30 | 1000 | A: 173 x 217<br>B: 139 x 217<br>C: 139 x 217<br>D: 91 x 217 | 151 | Total:<br>8 hr 57m 55s |
| 4 | | Tidy up | 55 | FIB | 30 | 300 | A: 97 x 6<br>B: 98 x 7<br>C: 107 x 8<br>D: 73 x 9 | 75 | 2 rounds<br>~ 1 hr total |
| | | | 55 | FIB | 30 | 100 | A: 81 x 3<br>B: 83 x 8<br>C: 86 x 5<br>D: 60 x 5 | 50 | 42m |
| 5 | Undercuts | Undercut 1 | 0 | FIB | 30 | 300 | (1) 116 x 14<br>(2) 9 x 93 | 100 | Stop when cut through |
| 6 | | Undercut 2 | 0 | FIB | 30 | 300 | (1) 119 x 19 (2) 7 x 68 | 100 | Stop when cut through |
| 7 | Lift out | needle attach | 0 | FIB | 30 | 250 | 7 x 10 | Until attached | ~11m |
| 8 | | cut tab | 0 | FIB | 30 | 300 | 10 x 156 | 100 | Until cut<br>(3-4m) |
| 9 | Attach | Attach to grid | 0 | FIB | 15 | 120 | ~80 x 20 | 5 | Stop when attached/no gap |
| 10 | | Needle cut | 0 | FIB | 30 | 100 | 23 x 1 | Until cut | <2m |
| 11 | Final prep (optional) | Extra dep | 55 | FIB | 30 | 30 | | | |
| 12 | | Tidy up/polish | 55 | FIB | 30 | 100 | | | |

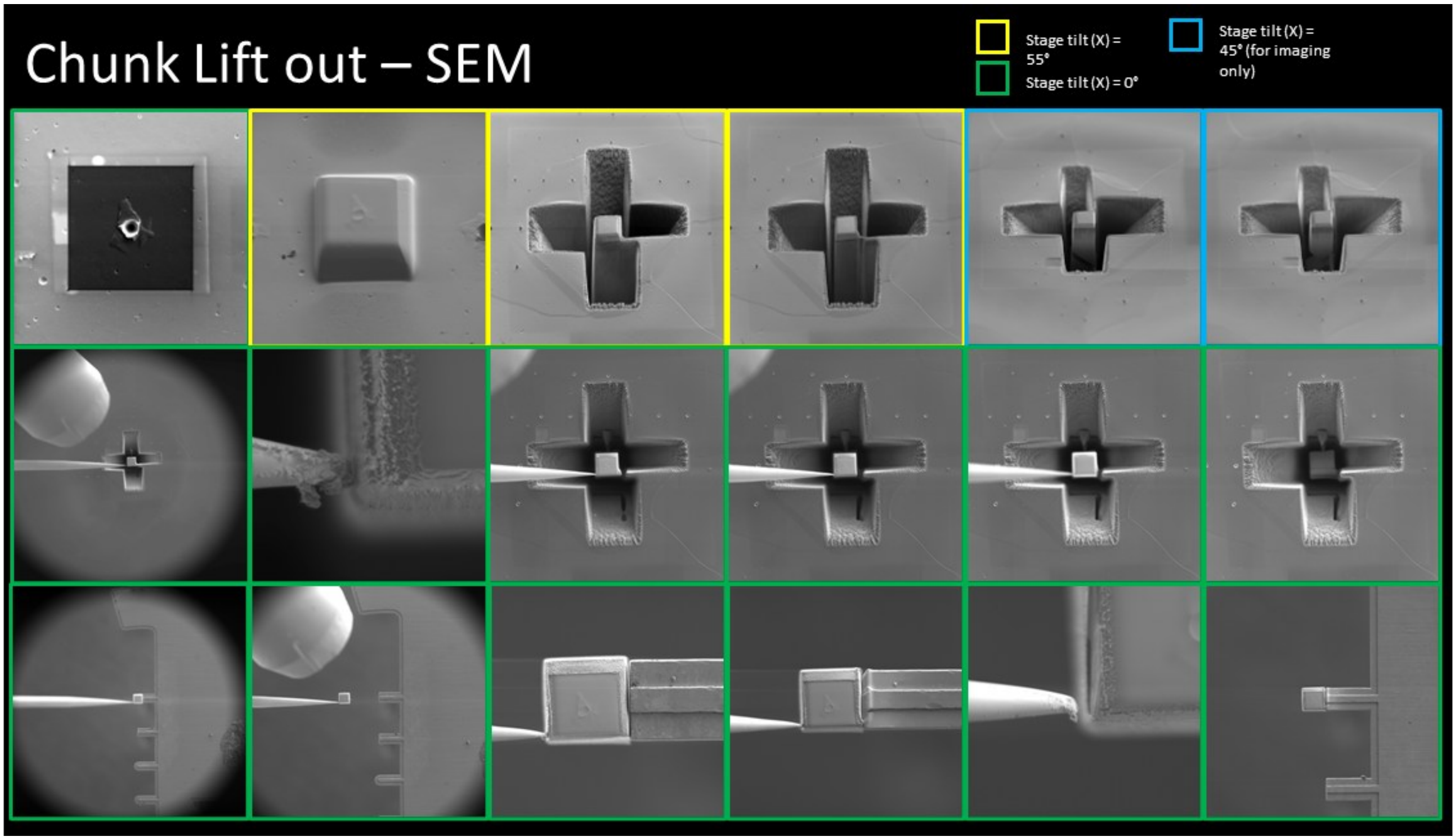

*Figure S1 – Chunk lift out steps compilation: e-beam deposition to attachment to grid – SEM images*

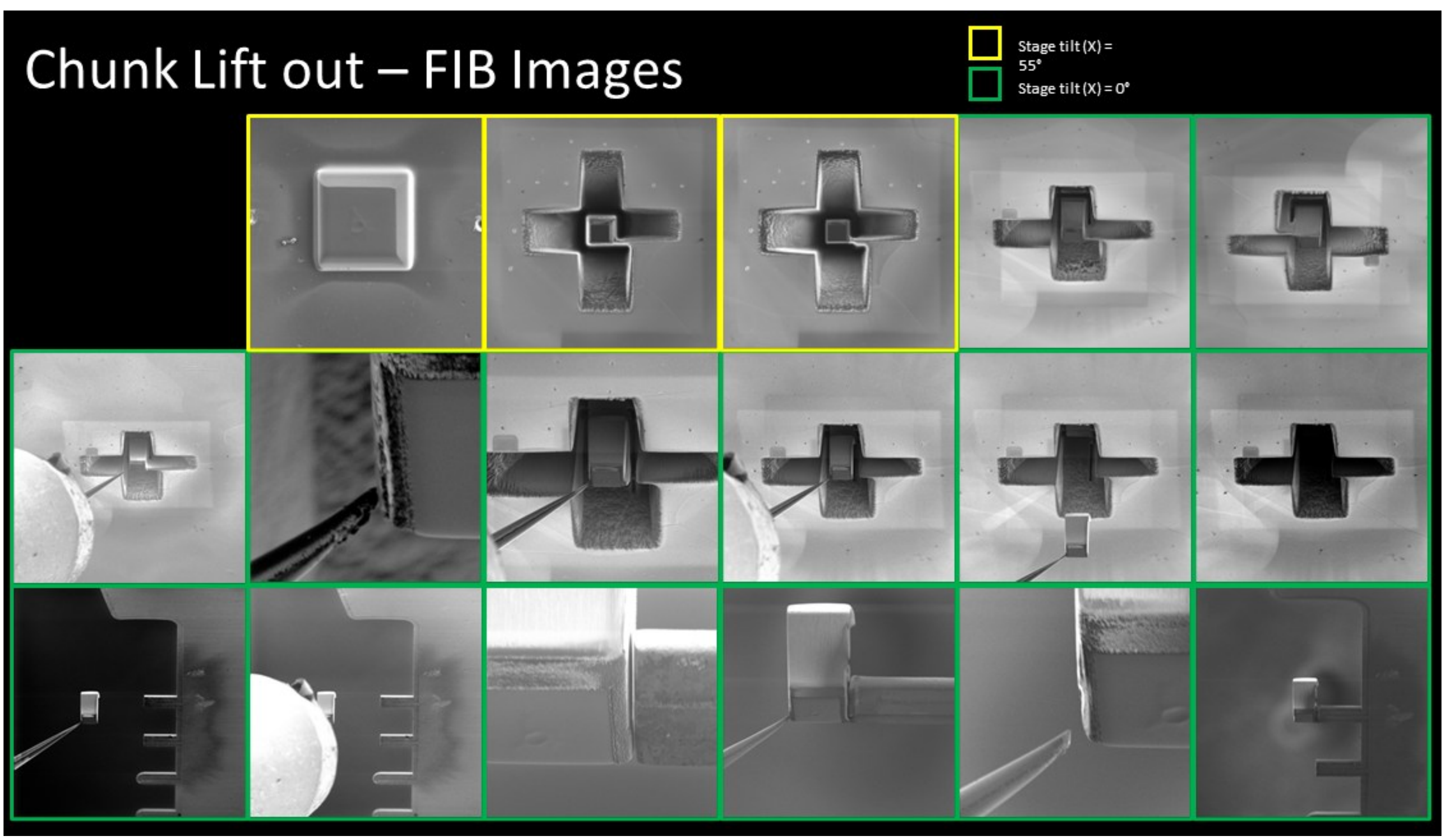

*Figure S2 – Chunk lift out steps compilation: e-beam deposition to attachment to grid – FIB images*